\documentclass[aps,floatfix,twocolumn]{revtex4}
\usepackage{graphics,epsfig}
\usepackage{graphicx}

\begin{document}

\title{Energy conditions and a bounce in FLRW cosmologies}
\author{Benjamin K. Tippett and Kayll Lake \cite{email}}
\affiliation{Department of Physics, Queen's University, Kingston,
Ontario, Canada, K7L 3N6 }
\date{\today}

\begin{abstract}
Exploiting the existence of two ``cosmological" constants, one
associated with the classical Lovelock theorem and one with the
vacuum energy density, we argue, in a model independent way, that
in spatially closed FLRW cosmologies with a positive definite
effective cosmological constant there exists a range in this
constant that serves as a sufficient condition for the
satisfaction of the null, weak, strong  and dominant energy
conditions at a bounce. The application of energy conditions is
not unambiguous and we show how the bounce can be considered
classically and how, we believe more reasonably, it can be
considered a matter of quantum cosmology.
\end{abstract}

\maketitle

\textit{Introduction.} The idea of a bounce in
Friedmann-Lema\^{i}tre-Robertson-Walker (FLRW) cosmologies is an
idea that is as old as spacetime cosmology itself \cite{old} but
also one with current appeal \cite{current} motivated primarily by
M-theory, branes and extra-dimensions \cite{Steinhardt}. A bounce,
as opposed to a big bang, obviously meets with many problems when
trying to reproduce the universe we observe. Moreover, there is a
widely held belief that a bounce must violate classical energy
conditions \cite{visser1}. However, the application of classical
energy conditions is not unambiguous and here we exploit the
existence of two ``cosmological" constants, one associated with
the classical Lovelock theorem and one with the vacuum energy
density, and argue, in a model independent way, that in spatially
closed FLRW cosmologies with a positive definite effective
cosmological constant (as explained precisely below) there exists
a range in this constant that serves as a sufficient condition for
the satisfaction of the null, weak, strong  and dominant energy
conditions at a bounce. Further, depending on how the energy
conditions are applied, we show how a bounce might be considered
classically and how, we believe more correctly, it must be
considered a matter of quantum cosmology. In this analysis we
offer no explanation as to why there should be a bounce but rather
only examine the consequences of a bounce as regards the
application of energy conditions.

 The suggestion that we live in a
spatially closed universe with a positive definite observed
cosmological constant is in accord with the recent Hubble Space
Telescope observations of Riess \textit{et al.} \cite{Riess},
observations that not only confirm earlier reports that we live in
an accelerating universe \cite{Riess98} \cite{Perlmutter}, but
also sample the transition from deceleration to acceleration. In
contrast, observations with the Wilkinson Microwave Anisotropy
Probe \cite{wmap} argue strongly that the FLRW cosmology is
spatially flat  \cite{lake1}. In the spatially flat case we argue
that a bounce necessarily violates energy conditions no matter how
they are applied. Because our application of the energy conditions
differs in a significant way from previous works, we start our
discussion from first principles.

\bigskip
\maketitle \textit{Background Geometry.} The Cosmological
Principle is the working hypothesis that the Universe is described
by a spacetime that is spatially isotropic globally. It follows
from the theorems of Robertson and of Walker \cite{rw} that the
spacetime is represented by by the
Friedmann-Lema\^{i}tre-Robertson-Walker (FLRW) metric, which, up
to a coordinate transformation, is given by \cite{conventions}
\begin{equation}
ds^2=a^2d\bar{s}_{\Gamma}^2-dt^2 \label{rw}
\end{equation}
where $a=a(t)$, normally $\in C^2$, and $d\bar{s}_{\Gamma}^2$ is
the metric of a three space of constant curvature. In  $\Gamma$ we
scale the coordinates so that $k \equiv R_{a}^{b}R_{b}^{a}/2
(R_{a}^{a})^2 = \pm1,0$ where $R_{a}^{b}$ is the Ricci tensor of
$\Gamma$.

\bigskip

The metric (\ref{rw}) has only two independent (scalar) invariants
derivable from the Riemann tensor without differentiation
\cite{invariants}. These are usually taken to be the Ricci scalar
$R$ ($\equiv R_{\alpha}^{\alpha}$, $R_{\alpha}^{\beta}$ the Ricci
tensor of the spacetime) and first Ricci invariant $R1$ ($\equiv
S_{\alpha}^{\beta}S_{\beta}^{\alpha}/4 \equiv S^2$,
$S_{\alpha}^{\beta}$ the trace-free Ricci tensor $\equiv
R_{\alpha}^{\beta}-\delta_{\alpha}^{\beta}R/4$). Without any loss
in generality we can also take the two independent invariants to
be $\rho$ and $p$ defined as follows:
\begin{equation}
8 \pi \rho \equiv R/4+sign(S)\sqrt{3 S^2} -\mathcal{C} \label{rho}
\end{equation}
and
\begin{equation}
8 \pi p \equiv -R/4+sign(S)\sqrt{S^2/3} +\mathcal{C},  \label{p}
\end{equation}
where $\mathcal{C}$ is a constant. In the coordinates of
(\ref{rw}) these evaluate to
\begin{equation}
8 \pi \rho = \frac{3}{a^2}(k+\dot{a}^2) -\mathcal{C},
\label{rhoeval}
\end{equation}
and
\begin{equation}
8 \pi p = -\frac{1}{a^2}(k+\dot{a}^2+2a\ddot{a}) +\mathcal{C},
\label{peval}
\end{equation}
respectively where $^{.}\equiv d/d t$. Since $\rho$ and $p$ are
invariants, they distinguish scalar polynomial singularities of
the metric (\ref{rw}) directly \cite{kretschmann}.

\bigskip
\textit{Physical Interpretation.} A remarkable feature of four
dimensions is the fact that the only divergence free two index
tensor $A^{\alpha}_{\beta}$ derivable from the metric tensor and
its first two derivatives is (up to a disposable multiplicative
constant) \cite{lovelock}
\begin{equation}
A^{\alpha}_{\beta}=G_{\alpha}^{\beta}+ \lambda_{c}
\delta_{\alpha}^{\beta}, \label{lovelock}
\end{equation}
where $G_{\alpha}^{\beta}$ is the Einstein tensor ($\equiv
R_{\alpha}^{\beta}-\frac{1}{2}R \delta_{\alpha}^{\beta}$) and
$\lambda_{c} $ is a constant. We write Einstein's equations in the
form
\begin{equation}
G_{\alpha}^{\beta}+\lambda_{c} \delta_{\alpha}^{\beta}=8 \pi
_{M}T_{\alpha}^{\beta} + 8 \pi _{V}T_{\alpha}^{\beta},
\label{einstein}
\end{equation}
where $_{M}T_{\alpha}^{\beta}$ signifies the matter contribution
to (\ref{rw}) and $_{V}T_{\alpha}^{\beta}$ the vacuum
contribution. Here we simply assume that the latter is of the
usual form $8 \pi_{V}T_{\alpha}^{\beta} \equiv \lambda_{v}
\delta_{\alpha}^{\beta}$ where $\lambda_{v}$ a constant. Central
to our considerations here is the treatment of the constant
$\lambda_{c}$ which, it is fair to say, is usually ignored.

\bigskip

It is useful at this point, prior to the application of energy
conditions, to review three distinct ways of proceeding with
(\ref{einstein}) \cite{lake2}. Because of the form of (\ref{rw})
we can always consider the source a perfect mathematical fluid
(including, for example, scalar fields \cite{ellis}) with
associated streamlines comoving with $\Gamma$ \cite{unique}. The
flow (congruence of unit timelike vectors $u^{\alpha}$) and normal
field $n^{\alpha}$ (in the tangent space of the associated
Lorentzian two-space) are uniquely determined. The energy density
and isotropic pressure (including bulk viscosity) are now defined
by
\begin{equation}
\rho \; \equiv \; T_{\alpha}^{\beta}
u^{\alpha}u_{\beta}\label{rhodef}
\end{equation}
and
\begin{equation}
p \; \equiv \; T_{\alpha}^{\beta} n^{\alpha}n_{\beta}\label{pdef}
\end{equation}
respectively where $T_{\alpha}^{\beta}$ is the energy-momentum
tensor of the mathematical fluid. The definitions (\ref{rhodef})
and (\ref{pdef}) reproduce (\ref{rhoeval}) and (\ref{peval}) (now
recognized as the Friedmann equations \cite{friedmann}) from the
field equations
\begin{equation}
G_{\alpha}^{\beta}+\mathcal{C} \delta_{\alpha}^{\beta}=8 \pi
T_{\alpha}^{\beta}. \label{einsteingeneral}
\end{equation}
The question of interest here is the treatment of the constants
$\lambda_{c}$ and $\lambda_{v}$ in the procedure of applying the
energy conditions. The three distinct ways of proceeding are:

\bigskip
(i) $\mathcal{C}=\lambda_{c}-\lambda_{v} \equiv \Lambda$: First
note that $\Lambda$ is an observable (let us call it the observed
cosmological constant). The apparent smallness of $\Lambda$ arises
out of the almost perfect cancellation of $\lambda_{c}$ and
$\lambda_{v}$ an explanation of which constitutes a standard form
of one of the ``cosmological constant problems" \cite{weinberg}.
The procedure of setting $\mathcal{C}=\lambda_{c}-\lambda_{v}
\equiv \Lambda$ is the common one when energy conditions are not
under consideration and, for example, is the procedure used
(implicitly) to construct the observers
$\Omega_{\Lambda}-\Omega_{M}$ plane \cite{lake1}. As regards the
application of energy conditions, this procedure is \textit{ad
hoc} since the vacuum contribution is, without explanation,
extracted from the energy-momentum tensor and considered part of
the geometry. We write the associated energy density and isotropic
pressure of the perfect fluid source as $\rho$ and $p$
respectively. These are given by (\ref{rhoeval}) and
(\ref{peval}). (We view $\mathcal{C}=-\lambda_{v}$ as a special
case where one simply assumes, without explanation, that
$\lambda_{c}=0$.)

\bigskip
(ii) $\mathcal{C}=\lambda_{c}$: In this procedure the vacuum
contribution is not extracted from the energy-momentum tensor but
rather considered part of it. The constant $\lambda_{c}$ is not
directly observable. Rather, its value can only be inferred from
the fact that $\lambda_{c} \sim \lambda_{v}$ and with
$\lambda_{v}$ calculated as in the standard model. The perfect
fluid source now has an associated energy density and isotropic
pressure given by $\tilde{\rho} \equiv \rho-\frac{\lambda_{v}}{8
\pi}$ and $\tilde{p} \equiv p+\frac{\lambda_{v}}{8 \pi}$
respectively. These are given by (\ref{rhoeval}) and (\ref{peval})
with $\rho$ and $p$ replaced by $\tilde{\rho}$ and $\tilde{p}$.
This procedure, as regards the application of energy conditions,
is clearly more natural than (i).

\bigskip
(iii) $\mathcal{C}=0$: In this procedure the geometrical
contribution from $\lambda_{c}$ and the contribution from the
vacuum $\lambda_{v}$ are both considered part of the
energy-momentum tensor.  Exactly why the geometrical contribution
$\lambda_{c}$ should be considered part of energy-momentum tensor
is left unexplained. The perfect fluid source now has an
associated energy density and isotropic pressure given by
$\bar{\rho} \equiv \rho-\frac{\lambda_{v}-\lambda_{c}}{8 \pi}$ and
$\bar{p} \equiv p+\frac{\lambda_{v}-\lambda_{c}}{8 \pi}$
respectively. Again these are given by (\ref{rhoeval}) and
(\ref{peval}) with $\rho$ and $p$ replaced by $\bar{\rho}$ and
$\bar{p}$. (This procedure can also be considered as a special
case of (ii) where one simply assumes, again without explanation,
that $\lambda_{c}=0$.)

\bigskip
\textit{Energy Conditions.} We now impose standard energy
conditions on the mathematical fluid \cite{he} \cite{visser}
\cite{poisson}. The local energy conditions considered here are
the null energy condition (NEC), weak energy condition (WEC),
strong energy condition (SEC) and dominant energy condition (DEC)
\cite{clarification}. We let EC designate the simultaneous
satisfaction of all four energy conditions. In the present context
we are always dealing with a perfect fluid and in that case the
conditions are given in case (i) by
\begin{equation}
\hbox{NEC} \iff \quad (\rho + p \geq 0 ),
\end{equation}
\begin{equation}
\hbox{WEC} \iff \quad (\rho \geq 0 ) \hbox{ and } (\rho + p \geq
0),
\end{equation}
\begin{equation}
\hbox{SEC} \iff \quad (\rho + 3 p \geq 0 ) \hbox{ and } (\rho + p
\geq 0),
\end{equation}
and
\begin{equation}
\hbox{DEC} \iff \quad (\rho \geq 0 ) \hbox{ and } (\rho \pm p \geq
0),
\end{equation}
where $\rho$ and $p$ are replaced by $\tilde{\rho}$ and
$\tilde{p}$ in case (ii) and by $\bar{\rho}$ and $\bar{p}$ in case
(iii). From (\ref{rhoeval}) and (\ref{peval}) for case (i) we find
\begin{equation}
\rho+3p \geq 0 \iff \mathcal{C} \geq \frac{3 \ddot{a}}{a},
\end{equation}
\begin{equation}
\rho+p \geq 0 \iff  k+\dot{a}^2 \geq a \ddot{a}, \label{rhop}
\end{equation}
\begin{equation}
\rho-p \geq 0 \iff \mathcal{C} \leq \frac{2}{a^2}(k +
\dot{a}^2)+\frac{\ddot{a}}{a},
\end{equation}
and
\begin{equation}
\rho \geq 0 \iff \mathcal{C}\leq \frac{3}{a^2}(k + \dot{a}^2),
\end{equation}
again with $\rho$ and $p$ are replaced by $\tilde{\rho}$ and
$\tilde{p}$ in case (ii) and by $\bar{\rho}$ and $\bar{p}$ in case
(iii). Case (iii) has been considered in detail previously
\cite{visser1}. As we show below, this case is not of interest for
our present considerations. From the foregoing conditions it
follows that \cite{note1}
\begin{equation}
\frac{3 \ddot{a}}{a}\leq \mathcal{C} \leq \frac{2}{a^2}(k +
\dot{a}^2)+\frac{\ddot{a}}{a} \Rightarrow \hbox{EC}
\label{sufficient}.
\end{equation}

\bigskip
\textit{A bounce.} We characterize regular minima via the
conditions
\begin{equation}
\dot{a}=0,\;\;\;\ddot{a}>0,\;\;\;a>0 \label{regularminima}
\end{equation}
at, say, $t=t_{m}$. It follows immediately from (\ref{rhop}) that
$k=1$ at a regular minimum (and in particular, $k=0$ is not a
possibility). Whereas the timelike convergence condition
($R^{\alpha}_{\beta}W_{\alpha}W^{\beta} \geq 0 \;\; \forall$
timelike $W^{\alpha}$) fails at all regular minima (e.g.
$W^{\alpha}=u^{\alpha}$), it follows from (\ref{sufficient}) that
there remains in general a non-vanishing interval in $\mathcal{C}$
sufficient for the satisfaction of the energy conditions in cases
(i) with $\lambda_{c}>\lambda_{v}$ and (ii) with $\lambda_{c}>0$.
Indeed with $k=1$ and (\ref{regularminima}) it follows that the
range in $\mathcal{C}$ can be narrowed to give
\begin{equation}
\frac{3 \ddot{a}}{a}\leq \mathcal{C}\leq \frac{2}{a^2} \Rightarrow
\hbox{EC}. \label{sufficient1}
\end{equation}
The marginal case $\dot{a}=\ddot{a}=0,\;\;a>0$ distinguishes
flatness at $t=t_{m}$ ($k=\mathcal{C}=0$) and $k=1$ with the
obvious substitutions in (\ref{sufficient}) and
(\ref{sufficient1}). In both these cases the timelike convergence
condition holds \cite{maxima}.

\bigskip
\textit{Discussion.} We have shown, in a model independent way,
that in a spatially closed FLRW universe there exits a range in
$\mathcal{C}>0$ in cases (i) and (ii) that serves as a sufficient
condition for the satisfaction of the standard energy conditions
at a bounce. Since we have used no properties of the scale factor
$a$, other than the existence of a bounce, nothing can be said
about the lower limit
\begin{equation}
\mathcal{C} \geq \frac{3 \ddot{a}}{a}|_{_{t_{m}}}>0.
\label{lowerlimit}
\end{equation}
Expressing the upper limits in terms of $a(t_{m})$ we have
\begin{equation}
a(t_{m}) \leq \sqrt{\frac{2}{\Lambda}},\;\;\;\; a(t_{m}) \leq
\sqrt{\frac{2}{\lambda_{c}}} \label{upperlimit}
\end{equation}
in cases (i) and (ii) respectively. In case (i), since $\Lambda <
\; \sim 10^{-56} cm^{-2}$ the lower bound on the upper limit to
$a(t_{m})$ is of the order of $4.5\;\; 10^3$ Mpc essentially
equivalent to the current Hubble scale ($\frac{c}{H_{0}}$) and so
one might hope that the existence or not of a bounce might be
answerable without recourse to quantum cosmology. The fact that
the energy conditions need not be violated at a bounce for
cosmologies consistent with the Hubble Space Telescope
observations is certainly of interest. However, as pointed out
above, this procedure, as regards the application of energy
conditions, is \textit{ad hoc} in that it extracts $\lambda_{v}$
from the energy-momentum tensor without any explanation.

In case (ii) we must \textit{assume} that $\lambda_{c}>0$ for the
energy conditions not to be violated at a bounce. Moreover, in
contrast to case (i), if we trust the standard model all the way
to the Planck scale, then the lower bound on the upper limit to
$a(t_{m})$ is within an order or so of the Planck scale itself and
the existence or not of a bounce becomes a matter of quantum
cosmology. This procedure has to be considered the preferred one
since the role of $\lambda_{c}$ and of $\lambda_{v}$ is unaltered
by \textit{ad hoc} assumptions. Whereas the classical energy
conditions may not play a fundamental role in quantum cosmology,
the fact that they need not be violated at a bounce is certainly
of interest. Moreover, the the avoidance of energy condition
violations at a bounce is tied to the mysterious cosmological
constant problem wherein the purely classical statement
(\ref{lovelock}) is, by the observation that $\lambda{_c} \sim
\lambda_{v}$, intimately connected to the non-classical
calculation of $\lambda_{v}$.

\begin{acknowledgments}
This work was supported by a grant to KL from the Natural Sciences
and Engineering Research Council of Canada.
\end{acknowledgments}


\begin{thebibliography}{}
\bibitem{conventions} We use a signature of
+2 and start with geometrical units. Latin indicies range from 1
to 3 and Greek indicies from 1 to 4.
\bibitem{kretschmann}In
particular, there is no need to look at the traditional
``Kretschmann" invariant $K=R_{\alpha \beta \gamma
\delta}R^{\alpha \beta \gamma \delta}$ ($R_{\alpha \beta \gamma
\delta}$ the Riemann tensor).
\bibitem{clarification} These conditions are usually stated as
inequatities that hold for any vector field of a stated type. By
this we understand that the inequalities must hold over the set of
all possible vector fields of a stated type. The inequalities are
in general not appropriate for a subset as the associated physics
can be entirely regular for the subset under inequalities that
differ from the stated conditions.
\bibitem{note1}Note that the existence of this interval (without
$\mathcal{C}$) is required if the spacetime is to satisfy the null
energy condition.
\bibitem{maxima} At regular maxima consider $\ddot{a}<0$ in the
foregoing and it follows that only for $k=1$ can $\mathcal{C}$ be
$>0$ in the sufficient condition for $\hbox{EC}$.
\end{thebibliography}
\end{document}